\documentclass[aps,twocolumn]{revtex4}
\usepackage{amssymb}
\usepackage{amsmath}
\usepackage{graphicx}

\begin{document}

\title{The non-analytic momentum dependence of spin susceptibility of Heisenberg
magnets in paramagnetic phase 
and its effect on critical exponents}
\author{A. A. Katanin}
\affiliation{Center for Photonics and 2D Materials, Moscow Institute of Physics and Technology, Institutsky lane 9, Dolgoprudny,
141700, Moscow region, Russia\\
M. N. Mikheev Institute of Metal Physics, Kovalevskaya str. 18, 620219,
Ekaterinburg, Russia.}

\begin{abstract}
We 
study momentum dependence of static magnetic susceptibility $\chi(q)$  in paramagnetic phase of Heisenberg
magnets and its relation to critical behavior within nonlinear sigma model (NLSM) at arbitrary dimension $2<d<4$.
In the first order of $1/N$ expansion, where $N$ is the number of spin components, 
we find $\chi(q)\propto[q^{2}+\xi
^{-2}(1+f(q\xi ))]^{-1+\eta /2}$, where $\xi $ is the correlation length, $q$ is the momentum, measured from magnetic wave vector, the universal scaling function $f(x)$ describes deviation from the standard Landau-Ginzburg momentum dependence. In agreement with previous studies at large $x$ we find $f(x\gg 1)\simeq (2B_{4}/N)x^{4-d}$;
the absolute value of the coefficient $B_4$ increases with $d$ at $d>5/2$. 
Using NLSM, we obtain the 
contribution of the``anomalous" term $\xi^{-2}f(q\xi )$ to the critical exponent $\nu $,
comparing it to the contribution of the non-analytical dependence,
originating from the 
critical exponent $\eta $ (the obtained
critical exponents $\nu $ and $\eta $ agree with previous studies). In the
range $3\leq d<4$ we find that the former contribution dominates, and fully
determines $1/N$ correction to the critical exponent $\nu $ in the limit $%
d\rightarrow 4.$ 
%The coefficient $B_4$ is found to vanish at $d=5/2$, which is related to
%the crossover from weak to strong spinon confiniement.   %By comparing the results to the earlier one for $%
%CP^{M-1}$ model, the anomalous contribution is associated with the spinon
%confinement, which becomes progressively stronger with increasing
%dimensionality of the system.
\end{abstract}

\maketitle

%\pacs{75.10.-b, 75.10.Jm, 75.40.Cx}

\section{Introduction}

The spatial or momentum dependence of response functions plays important
role in physical properties. The energy, corresponding to the spatial
dependence of the order parameter field $\mathbf{n}_{\mathbf{r}}$, weakly
changing in space, is 
%described by the contribution, 
proportional to $%
(\nabla \mathbf{n}_{\mathbf{r}})^{2}$ (as in the Ginzburg-Landau theory \cite%
{GL}). This 
%dependence 
yields at the critical point $S_{R}\sim R^{-(d-2)}$
decay of correlation function $S_{R}=\langle n_{0}^{\alpha }n_{\mathbf{R}%
}^{\alpha }\rangle $ with the distance $R,$ with $d$ being the
dimensionality of the system. At the same time, the interaction results in appearance of
anomalous critical exponent $\eta ,$ which determines the long-range
behavior of correlation functions $R^{-(d-2+\eta )}$ (see, e.g., Ref. \cite%
{Ma}). The exponent $\eta $ can vary from rather small value for 3D
Heisenberg model ($\eta \approx 0.04$) to substantial value for 2D Ising
model ($\eta =1/4$); substantial values $\eta =0.2-0.4$ were recently also
obtained for deconfined spinon theories  \cite%
{etadeconf,etalargeN}.

The scaling considerations away from the critical point predict spatial
dependence of correlation function $S_{R}=R^{-(d-2+\eta )}f(R/\xi )$ where $%
f(x)$ is some function, $\xi $ is correlation length. In momentum space the
corresponding dependence reads $S_{q}=q^{-2+\eta }g(q\xi )$. Simplest
function which fulfills this form is $S_{q}=A/(q^{2}+\xi ^{-2})^{1-\eta
/2}$, cf. Ref. \cite{Fisher}. This dependence generalizes Ornstein-Zernike result to include
anomalous exponent $\eta .$

However, the Ornstein-Zernike form (even with the exponent $\eta$) was argued to be not sufficient to explain experimental data. In this respect, non-analytic subleading corrections to scaling functions were proposed within Fisher-Langer theory \cite{FL,FL1} to explain anomalies of resistivity of transition metals near the magnetic phase transition \cite{Exp1,Exp2,Exp3}. These corrections were also invoked to explain peculiarities of the density-density correlation function near gas-liquid critical point \cite{Exp4,Exp5}. 

Theoretically, the corrections to correlation functions were 
obtained \cite{Aharony,Aharony1,Brezin,Wegner,Theumann,Nelson,Stutzer,Vicari} in large momentum $q$ limit 
%also in paramagnetic phase 
 within the linear sigma model (LSM). In case of specific heat critical exponent $\alpha<0$, the corresponding leading non-analytical term in the spin correlation function
%, obtained within the LSM, at large momenta 
reads $S_q\propto q^{-2+\eta-1/\nu}$  \cite{Aharony,Aharony1,Brezin,Wegner,Theumann,Nelson,Stutzer,Vicari}, where $\nu$ is the critical exponent of correlation length. This momentum dependence implies that 
the magnon self-energy, defined by $S_{q}=A/(q^{2}+\overline{\Sigma} _{q}+\xi ^{-2})^{1-\eta/2},$ acquires the non-analytic contribution $\overline{\Sigma} _{\mathbf{q}}\sim q^{2-1/\nu}$. This result can be also confirmed by renormalization group (RG) approach of Ref. \cite{2+e,2+e1,Chakravarty} 
in $d=2+\varepsilon $ dimensions, where $\nu=1/(d-2)+O(1)$, and therefore $\overline{\Sigma} _{\mathbf{q}}\sim q^{4-d}$.
The RG analysis \cite{Chakravarty,Sushkov} and $1/N$ expansion \cite%
{Chubukov} of $d=2$ Heisenberg magnets also agree with the above result for the self-energy, since they yield %, where 
% also show that the bosonic self-energy, defined by $%
% acquires 
the non-analytical momentum
dependence 
$\overline{\Sigma} _{\mathbf{q}}\sim q^{2}\ln ^{-1/(N-2)}(q\xi )$ 
at $q\gg
\xi ^{-1}$  ($N$ is the number of spin components), 
%also in agreement with the above mentioned result, since 
and $1/\nu\rightarrow 0$ in two
dimensions (the critical exponent $\eta =0$ for $d=2$, $N>2$). 

For the number of order parameter components $N>1$, the momentum dependence  $\overline{\Sigma} _{\mathbf{q}}\sim q^{4-d}$, discussed above for $d$ close to $2$, is identical to that, obtained from the longitudinal correlation function deeply in the ordered phase \cite{PP}. The latter dependence is produced by pair of spinons, and, therefore, reflects spinon deconfinement in the presence of long-range magnetic order \cite{Senthil,ChubukovStarykh,OurCP}. On approaching magnetic transition temperature $q^2$ term in the inverse Green's function becomes progressively more important (see, e.g., Refs. \cite{Our1N,MyLongTr}). This stresses possible relation of the non-analytical terms to the spinon (de)confinement. The non-analytical terms, obtained within LSM, also remind non-analytic contributions
to the spin susceptibility $\chi _{q}\sim q^{d-1}$ in itinerant systems \cite%
{Belitz,Maslov}.
%According to Refs.  similar contributions are also present at large $q$ in paramagnetic phase. 

%These studies point
%to the possibility of having additional non-analytic contributions to
%magnon propagator apart from those related to the anomalous exponent $\eta. $ 
% one expects the self-energy $\Sigma _{%
%\mathbf{q}}\sim q^{2-\varepsilon }$, similarly to the above mentioned result in the ordered phase. 

Previous theoretical studies of momentum dependence of susceptibility of $d>2$ Heisenberg model concentrated mainly on the large momentum asymptotics $q\gg\xi^{-1}$ of correlation functions and the interpolation formulae between Ornstein-Zernike and non-analytic dependences \cite{Wcshebor}. To study the universal properties of the Heisenberg model, in particular momentum dependence of correlation functions in the long-wave length limit, 
this model can be mapped to the non-linear sigma model (NLSM). The classical version of this model describes well the thermodynamic and statistical properties of Heisenberg magnets at finite not too low temperature\cite{Chakravarty,Chubukov,Our1N,OurRG,2+e1,2+e}. 
%At the same time, previous study of non-analytic terms in the dimensions not close to $d=2$ \cite{Aharony} was performed in the linear sigma model. Although due to universality this model yields the same critical behavior, as for the NLSM, the latter model 
The NLSM has certain advantages over the linear sigma model, previously used to calculate asymptotics of correlation functions, since it is applicable outside the critical regime. Also, in NLSM the universal part of the magnon self-energy is directly related to the correlation length via constraint equation, reflecting fixed spin value. This allows us
to study the effect of non-analytical terms on the critical exponents. %obtained previously, 

%It was argued some time ago \cite{2+e_special} and confirmed recently within the functional renormalization group approach \cite {Jakubczyk} that 
%the critical exponents are 
In the present paper we consider derivation of nonanalytic contributions to
momentum dependence of spin susceptibility in paramagnetic phase of Heisenberg magnets within NLSM,  study in details their structure with varying dimensionality, and their
effect on the critical behavior. We 
%study general structure of the self-energy at large moment $q\gg \xi^{-1}$ and
%show that the self-energy
%contains indeed the term $\Sigma _{\mathbf{q}}\sim q^{4-d}$ and 
determine closed analytical expression for the coefficient of the leading non-analytical term $q^{4-d}$ in the self-energy to the first order in $1/N$ in arbitrary dimension $2<d<4.$ The absolute value of the coefficient of the anomalous term becomes larger with increase of the system
dimensionality $d$, which is related to stronger spinon confinement with increasing dimensionality.
%Although our results for the non-analytic term agree with previous study \cite{Aharony}, we find closed analytical expression for the coefficient at the leading non-analytic term, which was previously expressed through infinite series.
%This term is similar to that 
%obtained long time ago in the ordered phase \cite{PP}. 
%Due to presence of finite correlation length the obtained term is subleading with respect to the ``standard" contribution $q^2$ to the self energy, which reflects confinement of spinons.
  We also
argue that the non-analytic term yields substantial contribution to the critical
exponent $\nu ,$ and therefore, via scaling relations, all other critical
exponents, except the exponent $\eta ,$ which is shown to be independent of
the presence of the term. 

\section{1/N expansion in the non-linear sigma model}

%\subsection{General formalism}

We consider the classical non-linear $O(N)$ sigma model 
\begin{eqnarray}
Z[\mathbf{h}] &=&\int D\sigma D\lambda \exp \left\{ -\frac{1}{2t}\int d^{d}%
\mathbf{r}\left[ (\mathbf{\nabla }\mbox
{\boldmath $\sigma $})^{2}\right. \right.  \notag \\
&&\left. \left. +i\lambda (\sigma ^{2}-1)-2t\mathbf{h}%
\mbox
{\boldmath $\sigma $}\right] \right\} ,  \label{NLSM}
\end{eqnarray}%
where $\mbox
{\boldmath $\sigma $}(\mathbf{r)}$ by the $N$-component field, $d$ is the
space dimensionality, $t=T/\rho _{s}$ is the coupling constant, $\rho _{s}$
is the spin stiffness. The constraint condition $\sigma ^{2}=1$ is taken
into account by introducing the auxiliary field $\lambda (\mathbf{r}).$ To
calculate the correlation functions we also introduce the external
non-uniform magnetic field $\mathbf{h}(\mathbf{r}).$ The model (\ref{NLSM})
is applicable to classical and quantum ferro- and antiferromagnets at finite
temperatures (in quantum case the temperature should not be too low:\ $JS\xi
^{-1}\ll T,$ $J$ is the exchange integral and $S$ is the spin value, see
Refs. \cite{Chakravarty,Chubukov,Our1N,OurRG}). The applicability of the classical model (\ref{NLSM}) to quantum ferro- and antiferromagnets at finite not very low temperatures is related to the fact that quantum renormalizations at finite temperatures can be absorbed into the spin stiffness $\rho_s$.
%, choosing the corresponding cutoff in momentum space $\Lambda\sim T/(JS)$ instead of $\Lambda\sim 1$ for the classical model. 
The model (\ref{NLSM}) is also applicable
to the quantum antiferromagnets in the ground state, in which case $t\sim
1/S $, and $d$ is the space-time dimensionality.\ 

To study non-analytical terms in the self-energy, we use $1/N$ expansion, which is
performed in the standard way\cite{Aharony,Polyakov,Chubukov,ArovasBook}. In contrast to 
the (self-consistent) spin-wave theory \cite{SSWT} and $2+\epsilon$ renormalization group approach \cite{Chakravarty,2+e1,2+e} this method allows to study systems with dimensionality $d$ not close to 2ß at not too low temperatures. 
%in contrast to the  .
%, since the latter approaches rely on weak confinement of spinons. 
After
integrating over $\mbox
{\boldmath $\sigma $}$ the partition function takes the form 
\begin{eqnarray}
&&Z\,[\mathbf{h}]%
\begin{array}{c}
=%
\end{array}%
\int D\lambda \exp (S_{eff}[\lambda ,h])  \label{Zef} \\
&&S_{eff}[\lambda ,h]%
\begin{array}{c}
=%
\end{array}%
\frac{N}{2}\ln \det \widehat{G}+\frac{1}{2t}\text{Sp}(i\lambda )  \notag \\
&&\ +\frac{t}{2}\text{Sp}\left[ h\widehat{G}h\right] ,  \label{Sef}
\end{eqnarray}%
where 
\begin{equation}
\widehat{G}=[-\mathbf{\nabla }^{2}+i\lambda ]^{-1}.  \notag
\end{equation}%
Since $N$ enters (\ref{Zef}) only as a prefactor in the exponent, expanding
near the saddle point generates a series in $1/N$. Below we treat the
paramagnetic phase, where the value of $\lambda =\lambda _{0}$ at the saddle
point is determined by the sum rule (constraint) $\langle \sigma ^{2}\rangle
=1,$ which takes the form 
\begin{equation}
1=Nt\int \frac{d^{d}\mathbf{k}}{(2\pi )^{d}}G^{nn}({k}),
\label{Constr0}
\end{equation}%
where $n=1...N$ and we account that Green's function of the field $\sigma $,  
\begin{equation}
G^{nn'}({k})%
\begin{array}{c}
=%
\end{array}%
\frac{1}{tZ[0]}\left[ \frac{\partial ^{2}Z[h]}{\partial h^{n}(\mathbf{k}%
)\partial h^{n'}(\mathbf{-k})}\right] _{h=0},
\end{equation}%
depends only on $k=|\mathbf{k}|$ due to
rotational symmetry in the considering long-wavelength limit, $h(\mathbf{k})$ is the Fourier transform of $h(\mathbf{r})$. Note that only
diagonal elements $G^{nn'}$ are nonzero. We use the cutoff $k<\Lambda $ of
momentum integrations.
%, where $\Lambda \sim 1$ for classical or ground state
%quantum model and $\Lambda \sim T/(JS)$ for quantum model at finite $T$, see
%Refs. \cite{Chubukov,Our1N,OurRG}. 
The Green's function represents the
rescaled (staggered) spin susceptibility $\chi ^{nn'}({k})=(S^2/\rho _{s})G^{nn'}(%
{k})$ and may be expressed within $1/N$ expansion as 
\begin{equation}
G^{nn}(k)=\left[ k^{2}+\Sigma (k)+m^{2}\right] ^{-1},
\end{equation}%
where $\Sigma (k)$
is the bosonic self-energy, defined such that $\Sigma(0)=0$, and $m$ is the renormalized mass of spin
excitations to first order in $1/N.$ We split the mass as $%
m^{2}=m_{0}^{2}+\delta m^{2}$ where we define $m_{0}$ in such a way that it
absorbs all non-universal ($\Lambda $-dependent) contributions, except
logarithmic terms (the latter contribute to critical exponents and included,
as well as regular terms, in $\delta m^{2}$, see below). The terms included
in $m_{0}$ determine the value of magnetic phase transition temperature
(or critical coupling constant),
which is defined by vanishing $m_{0}$ (the quantity $\delta m^{2}$ vanish
simultaneously, see below).

The self-energy $\Sigma (k)$ in the first order of $1/N$ expansion is given
by \cite{Chubukov} 
\begin{equation}
\Sigma (k)=\frac{2}{N}\int \frac{d^{d}\mathbf{q}}{(2\pi )^{d}}\frac{G_{0}(%
|\mathbf{k}+\mathbf{q}|)-G_{0}(q)}{\Pi (q)}  \label{CE}
\end{equation}%
where 
\begin{equation}
\Pi (q)=\int \frac{d^{d}\mathbf{p}}{(2\pi )^{d}}G_{0}(p)G_{0}(|\mathbf{p}+%
\mathbf{q}|)  \label{Pi}
\end{equation}%
and $G_{0}(k)=(k^{2}+m_{0}^{2})^{-1}.$ To first order in $1/N$ the sum rule (%
\ref{Constr0}) takes the form 
\begin{eqnarray}
1 &=&Nt\int \frac{d^{d}\mathbf{k}}{(2\pi )^{d}}G_{0}(k)  \label{Constr1/N} \\
&&-Nt\int \frac{d^{d}\mathbf{k}}{(2\pi )^{d}}G_{0}^{2}(k)\left[ \Sigma
(k)+\delta m^{2}\right]  \notag
\end{eqnarray}

\section{Results}

\subsection{Three dimensions}

In three dimensions the polarization operator (\ref{Pi}) reads 
\begin{equation}
\Pi (q)=\frac{1}{4\pi q}\arctan \frac{q}{2m_{0}}.  \label{Pi3}
\end{equation}%
Using this expression, we find from Eq. (\ref{CE}) (see details in Appendix
A.1)%
\begin{equation}
\Sigma (k)=\eta k^{2}\ln \frac{\Lambda }{(k^{2}+m_{0}^{2})^{1/2}}+\frac{2}{N}%
m_{0}^{2}F(k/m_{0}),  \label{SE3}
\end{equation}%
where $\Lambda $ is the momentum cutoff, $\eta =8/(3\pi ^{2}N)$ is the
standard exponent determining correlation function decay to first-order in $%
1/N$ for 3D $O(N)$ model (cf. Refs. \cite{Hikami,Chubukov,Our1N}), and we
have introduced a universal function 
\begin{eqnarray}
F(x) &=&\frac{1}{\pi }\int\limits_{0}^{\infty }q^{2}dq\left\{ \left[ \frac{1%
}{2qx}\ln \frac{(x+q)^{2}+1}{(x-q)^{2}+1}-\frac{2}{q^{2}+1}\right] \right. 
\notag \\
&&\left. \times \frac{q}{\arctan (q/2)}-\frac{4x^{2}}{3\pi q^{3}}\theta (q-%
\sqrt{x^{2}+1})\right\} .  \label{Fx3}
\end{eqnarray}%
Evaluating asymptotics of this function, we find%
\begin{equation}
F(x)\simeq \left\{ 
\begin{array}{cc}
\frac{4x^{2}}{9\pi ^{2}}-\frac{2x}{\pi }-\frac{4}{\pi ^{4}}(16-\pi ^{2})\ln
x+1.10334 & x\gg 1 \\ 
-0.24553x^{2} & x\ll 1%
\end{array}%
\right. .  \label{Fxa}
\end{equation}%
One\ can see that at $k\gg m_{0}$ apart from quadratic term $Ak^{2}$ the
self-energy contains also subleading non-analytical terms, proportional to $%
k $ and $\ln (k/m_{0})$ with the coefficients, which agree with Ref. \cite{Aharony}, but expressed in terms of elementary functions. These terms are not related to the exponent $\eta ,$
introduced by the first term in Eq. (\ref{SE3}). The plot of the function $%
F(x)$ together with its asymptotes is shown in Fig. 1. Note that despite the
function $F(x)$ is not positively defined, it is quadratic at small $x$, and
therefore, the leading term $k^{2}$ in the propagator overcomes the negative
contribution in the second line of Eq. (\ref{Fxa}), and the whole spectrum
is positively defined at large $N$ (including $N=3$).

\begin{figure}[t]
\includegraphics[width=0.48\textwidth]{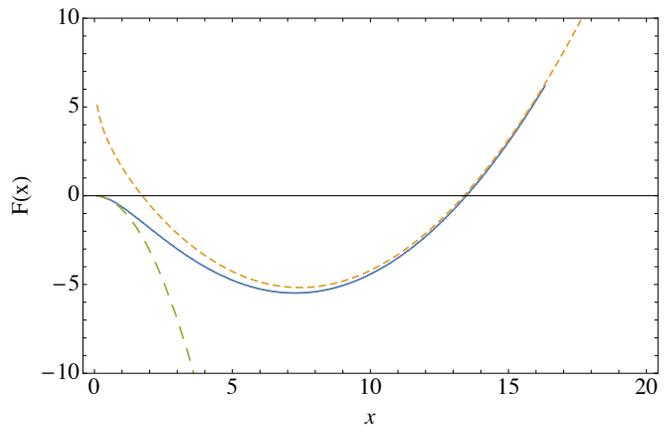}
\caption{(Color online) The plot of the function $F(x)$ (solid line) and its
large- and small-$x$ asymtotes (dashed lines). }
\end{figure}

The transition temperature (or critical coupling constant) to the first
order in $1/N$ is obtained from Eq. (\ref{Constr1/N}) by putting $%
m_{0}=\delta m=0$. We find%
\begin{equation}
t_{c}=\frac{2\pi ^{2}}{N\Lambda }\left( 1+\frac{32}{9\pi ^{2}N}\right) .
\label{tc3}
\end{equation}%
Details of calculation of the mass terms $m_{0}$ and $\delta m$ are
presented in Appendix A.1. The mass $m_{0}$ behaves near the phase transition
as (see Eq. (\ref{m03A})) 
\begin{equation}
m_{0}\propto \frac{1}{t_{c}}-\frac{1}{t}.  \label{m03}
\end{equation}%
%
%
%
%
%
%
%To obtain critical exponents, we need also singular correction to $\delta m.$
Summation of logarithmic contributions to the Eq. (\ref{Constr1/N}) yields 
\begin{equation}
\delta m^{2}=\delta m_{\eta }^{2}+\delta m_{k}^{2}+...=\left( 3\eta +\frac{16%
}{\pi ^{2}N}\right) m_{0}^{2}\ln \frac{\Lambda }{m_{0}}+...  \label{dm3}
\end{equation}%
where $\delta m_{\eta }^{2}\ $and $\delta m_{k}^{2}$ (as well as respective
terms in the right hand side) denote the contribution of the first
(proportional to $\eta $)\ term in the self-energy $\Sigma _{k}$, Eq. (\ref%
{SE3}), and linear in $k$ term in $\Sigma _{k}$, originating from the second
term in Eq. (\ref{SE3}) (the other terms in Eq. (\ref{Fxa}) apart from the
linear one do not contribute to the singular term in Eq. (\ref{dm3})); the
dots stand for the non-singular terms proportional to $m_{0}^{2}$.
Collecting all logarithmic contributions, which are of the order $1/N$, to
the Green's function and transforming them to the respective powers to
introduce $1/N\ $corrections to critical exponents, we obtain (see Appendix
A.1)%
\begin{equation}
G(k)=\frac{1}{\left\{k^{2}+\xi ^{-2}[1+f(k\xi)] \right\}^{1-\eta/2}}  \label{G3}
\end{equation}%
where $f(x)=(2/N)\{F(x)-(2/(3\pi ^{2}))\ln [1/(x^{2}+1)]\},$ $\xi
=m_{0}^{-\nu }\propto (t-t_{c})^{-\nu },$ is the correlation length, $\nu
=1-\eta -\nu _{k}$ is the corresponding critical exponent. The contribution $%
\eta $ originates again from the first term in the self-energy $\Sigma _{k},$
Eq. (\ref{SE3}), while$\ \nu _{k}=8/(\pi ^{2}N)$ originates from the linear
in $k$ term in $\Sigma _{k}$. Although the sum of the two terms yields the
standard result $\nu =1-32/(3\pi ^{2}N)\simeq 0.64$ ($N=3$) \cite%
{Hikami,Chubukov,Our1N}, our result allows to discriminate the contribution
of non-analytic terms originating from the anomalous exponent $\eta $ and
the linear in $k$ term in the self-energy. One can see that the latter is
three times larger than the former, i.e. main contribution to the $1/N$
correction to the critical exponent $\nu $ originates from the linear in $k$
term of the self-energy. Indeed, excluding at $N=3$ the term, related to $\eta $,
yields $\nu =0.73,$ but excluding linear in $k$ term we get $\nu =0.9$,
which is far from the $1/N$ result. This shows importance of non-analytic
contribution to the self-energy for critical exponents in three dimensions.

To estimate the deviation from the Ornstein-Zernike form we introduce the Green function $G_{\rm OZ}=1/(\xi^{-2}+\varkappa k^2)$, where the coefficient 
$\varkappa=1+8/(9\pi^2 N)$ takes into account renormalization of the coefficient at $k^2$ by the first-order $1/N$ expansion, see Eq. (\ref{Fxa}). The momentum dependence of the ratio of Green function (\ref{G3}) to the Ornstein-Zernike one is shown in Fig. 2. One can see that the obtained Green function $G(k)$ essentially differs from both, the Ornstein-Zernike $G_{\rm OZ}(k)$ and modified dependence $G_{\rm OZ}^{1-\eta/2}(k)$. In particular, in comparison to the  $G^{1-\eta/2}_{\rm OZ}(k)$ dependence two flection points appear. Interestingly, these flection points can be seen on the experimental data near liquid-gas critical point \cite{Exp5}, although the present theory, based on $1/N$ expansion, is not applicable directly to the $N=1$ case. 

\begin{figure}[t]
\includegraphics[width=0.48\textwidth]{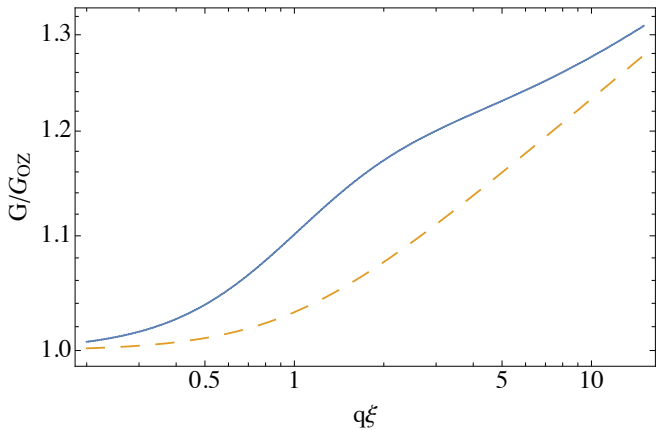}
\caption{(Color online) The dependence of the ratio of Green functions $G/G_{\rm OZ}$ on $q\xi$ at $N=3$ (solid line) in comparison to the correction $G_{\rm OZ}^{-\eta/2}$ according to the modified Ornstein-Zernike dependence (dashed line, see text). }
\end{figure}

\subsection{Arbitrary $2<d<4$}

Let us now generalize the obtained results to arbitrary dimension $2<d<4.$
Performing integration in Eq. (\ref{CE})\ we find in this case (see Appendix
A.2) 
\begin{equation}
\Sigma (k)=\eta k^{2}\ln \frac{\Lambda }{(k^{2}+m_{0}^{2})^{1/2}}+\frac{2}{N}%
m_{0}^{2}F_{d}(k/m_{0}),  \label{SEd}
\end{equation}%
with $\eta =-(2/N)\left( 4-d\right) \sin \left( \pi d/2\right) \Gamma
(d-1)/(\pi d\Gamma \left( d/2\right) ^{2})$ being the value of the exponent
for correlation function to first order in $1/N$ \cite{Hikami} and the
function $F_{d}(x)$ is given by the Eq. (\ref{Fd}). The expansion of this
function at $x\gg 1$ reads%
\begin{eqnarray}
F_{d}(x) &\overset{x\gg 1}{=}&B_{0}+B_{0}^{\prime }\ln
x+B_{2}x^{2}+B_{4}x^{4-d}+B_{6}x^{6-2d}  \notag \\
&+&B_{8}x^{8-3d}+...,  \label{Fdexp}
\end{eqnarray}%
where the coefficients at the $\ln x$ and at the highest power of $x$ (apart
from quadratic term) are given by%
\begin{eqnarray}
B_{0}^{\prime } &=&\frac{2(2d-5)\sin ^{2}(\pi d/2)\Gamma \left( 2-\frac{d}{2}%
\right) \Gamma (d-2)}{\pi ^{2}\Gamma \left( d/2\right) },  \notag \\
B_{4} &=&\frac{(5-2d)\Gamma (d-1)}{2\Gamma \left( d/2\right) ^{2}}.
\label{B04}
\end{eqnarray}%
First terms of the expansion (\ref{Fdexp}) were considered within the LSM in Ref. \cite{Aharony}; the obtained coefficients $B_{0}^{\prime } $ and $B_4$ coincide numerically with those obtained in LSM \cite{Note_Aharony1}, although here we present simple analytical expression for $B_4$ instead of the series, obtained in Ref. \cite{Aharony}.  
The plot of the dependence of $B_{4}$ on dimensionality $d$ is shown in Fig.
3. The coefficient $B_{4}$ decreases with increasing dimensionality and
becomes negative for $d>5/2$. We note the following peculiarities of the
function $F_{d}(x)$.

\begin{figure}[b]
\includegraphics[width=0.48\textwidth]{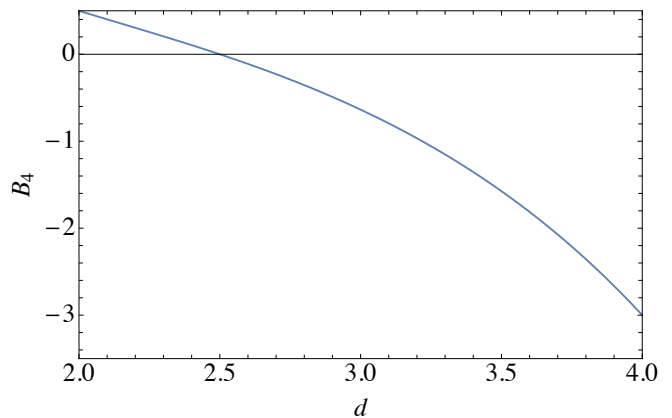}
\caption{(Color online) The dependence of the coefficient $B_4$ at the
leading power $k^{4-d}$ of the momentum dependence of the self-energy on the
dimensionality $d$}
\label{B4}
\end{figure}

(i) At $d\rightarrow 2$ we have $B_{0}^{\prime }=0,$ while all the powers of 
$x=k/m_{0}$ in Eq. (\ref{Fdexp})\ approach $2$. This yields the non-analytic
momentum dependence obtained in Ref. \cite{Chubukov}, $F_{2}(x)\sim x^{2}\ln
\ln x.$ Moreover, as it is argued in Ref. \cite{Chubukov} in this limit
summation of infinite series of $1/N\ $expansion is required, which yields $%
k^{2}(1-(1/N)\ln \ln (k/m_{0}))\rightarrow k^{2}\ln ^{-1/(N-2)}(k/m_{0})$.
Therefore, in the limit $d\rightarrow 2$ one can not restrict oneself to the
finite number of terms neither in the expansion (\ref{Fdexp}) nor in $1/N$
expansion (the latter problem can however be solved by replacement $%
N\rightarrow N-2$ in the lowest order $1/N$ term and transforming
logarithmic contributions into powers, cf. Ref. \cite{Chubukov}). 

It was
observed earlier that the case of $d\ $close to $2$ is described well by $%
d=2+\varepsilon $ expansion \cite{2+e1,2+e} and for $N=3$ by the $1/M$
expansion of non-compact $CP^{M-1}$ model \cite{OurCP}. Since the latter
model was argued to be applicable near deconfined quantum critical points 
\cite{deconf,etalargeN}, the limit $d\rightarrow 2$ can be viewed as
corresponding to the weakly confined spinons, cf. Ref. \cite{ChubukovStarykh}%
. Fully deconfined spinons are characterized by $k^{d-4}$ dependence of the
spin correlation function (obtained as a convolution of two spinon Green's
function with the dependence $1/q^{2}$ each) \cite{Senthil,ChubukovStarykh,OurCP}, similarly to the longitudinal correlation function in the ordered phase \cite{PP}. 
Therefore the obtained leading
non-analytical term in the Eq. (\ref{Fdexp}), $B_{4}x^{4-d}$, for $d$ close
to 2 can be considered as a ``trace" of this tendency to deconfinement. This
term, however, is not dominating over $k^{2}$ dependence, and the tendency
to confinement dominates. 
%We also note that the $k^{4-d}$ term in the self-energy in the ordered phase also becomes progressively less important on approaching magnetic transition temperature, and dominates only at small momenta $k\ll \overline{\sigma}^{2/(d-2)}$, where $\overline{\sigma}$ is the (sublattice) magnetization, as discussed in Appendix C of Ref. \cite{Our1N}. In the symmetric phase this would correspond to the region $k\ll m_0$, where the self-energy is however small in comparison to the bare term $m_0^2$, related to presence of finite correlation length.  

(ii) With decreasing dimensionality from $d=4$ at the set of dimensions 
%some peculiar values 
$d_i=2i/(i-1),$ where $i>2$ is an integer, i.e.
at $d=3,8/3,5/2,...,$ the term proportional to $B_{2i}$  in the expansion (\ref{Fdexp}) becomes relevant, since the corresponding power changes sign. The dimensions $d_i$ coincide with those, at which the operators $(\mbox{\boldmath $\phi $}^2)^i$  in LSM (which can be viewed as a soft constrain version of NLSM) become relevant and corresponding new fixed points in renormalization group flow appear. Since the fixed point structure of LSM and NLSM is expected to be the same, one can consider the non-analytical terms as related to these fixed points. One can verify that the corresponding coefficients $B_{2i}$ are
logarithmically singular in these special dimensions $d_i$ providing additional
contribution to $B_{0}^{\prime }.$ 
%This does not happen at arbitrary $d.$ 
From this point of view, the leading non-analytical term $k^{4-d}$ is always relevant for $d<4$ and it is related to the non-gaussian Wilson-Fisher fixed point, cf. Refs. \cite{Aharony1,Brezin,Wegner}. However, in contrast to the other coefficients $B_{2i}$ at the dimensions $d \rightarrow d_i$, the coefficient $B_4$ contains at $d\rightarrow 4$ the ratio of two logarithms $\ln(k/m)/\ln(\Lambda/k)$ (see Appendix A.3): the numerator reflects the logarithmic divergence of the integral in Eq. (\ref{CE}), while the denominator appears because of the logarithmic divergence of $\Pi(q)$ in four dimensions.  

(iii)\ For $d$ not too close to 2 only the leading terms $%
B_{2}x^{2}+B_{4}x^{4-d}$ are important, the latter provides non-analytical
contribution to the self-energy, which, as we will see below, yields
contribution to critical exponents, similarly to $d=3$ case. We note that
neither the results of $2+\varepsilon $ expansion, nor $1/M$ expansion of
non-compact $CP^{M-1}$ model become applicable for $d\gtrsim 3$ non-linear
sigma model (see, e.g., the discussion in Ref. \cite{OurCP}). This is in
line with the suggestion of Ref. \cite{2+e_special} (see also Ref. \cite{Jakubczyk}) that a sharp change of
critical exponents 
%with discontinuous derivatives on $d$ 
occurs somewhere in
the range $2<d<3$ and may imply stronger spinon confinement at $d\gtrsim 3$.

\begin{figure}[t]
\includegraphics[width=0.48\textwidth]{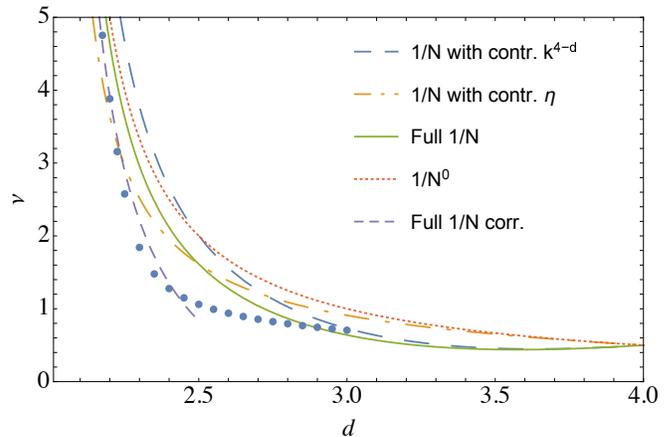}
\caption{(Color online) The dependence of the critical exponent $%
\protect\nu$ of $N=3$ non-linear sigma model on dimensionality $d$ in various approaches: zeroth order $1/N$ result (dotted line),
first order $1/N$ approach neglecting the exponent $\eta$ (long dashed line),
first order $1/N$ approach with only correction from the exponent $\protect%
\eta$ included (dash-dotted line), full first order $1/N$ resut (solid line),
and first order $1/N$ result with the replacement $N\rightarrow N-2$, which has to be performed for $d\rightarrow 2$ (short
dashed line). Dots show the results of functional renormalization group approach of Ref. \cite{Jakubczyk}.}
\label{model_with_peak}
\end{figure}

Following the same strategy, as for $d=3,$ we obtain the correction to the
magnetic transition temperature (or critical coupling constant) 
\begin{equation}
t_{c}=\frac{d-2}{NA_{d}\Lambda ^{d-2}}\left[ 1+\frac{\eta }{d-2}+B_{2}\right]
.  \label{tcd}
\end{equation}%
The calculation of the masses $m_{0}$ and $\delta m,$ as well as critical
exponent $\nu $ is performed in the same way as for $d=3$ and detailed in
Appendix A.2. While $m_{0}\propto (t-t_{c})^{1/(d-2)}$, the $1/N$ correction
to the mass reads%
\begin{equation}
\delta m^{2}=\frac{\eta d+(8B_{4}/\pi N)\sin \left( \pi d/2\right) }{d-2}%
m_{0}^{2}\ln \frac{\Lambda }{m_{0}}+...,  \label{dmd}
\end{equation}%
where the first and second terms in the numerator correspond to the
contribution of the first term in Eq. (\ref{SEd}) and $x^{4-d}$ term in the
asymptotic $F_{d}(x)$, the dots stand for the non-logarithmic terms. From
these contributions, we obtain 
\begin{equation}
\nu =\frac{1}{d-2}\left[ 1-\frac{\eta +(4B_{4}/\pi N)\sin \left( \pi
d/2\right) }{d-2}\right] .  \label{nud}
\end{equation}%
This critical exponent also coincides with the earlier known result of $1/N$
expansion \cite{Hikami}, but we again individuate here contributions of two
different effects: the anomalous exponent $\eta $ and $k^{4-d}$ term in the
self-energy. The contributions of these two effects to the critical exponent 
$\nu $ at $N=3$ and various dimensionality $d$ are plot in Fig. 4. One can see that
while for $d\rightarrow 2$ these two effects almost cancel each other, with
increasing dimensionality $d$ the $k^{4-d}$ term of the self-energy gives
progressively larger contribution; for $d\rightarrow 4,$ where the anomalous
exponent $\eta $ vanishes, the $k^{4-d}$ term gives the major contribution
to the $1/N$ correction to the critical exponent $\nu $. For completeness we
also show in Fig. 4 the result of the $1/N$ expansion with the replacement $%
N\rightarrow N-2,$ which, according Refs. \cite{Chubukov,OurCP} should be
performed in the limit $d\rightarrow 2$, allowing, e.g., to achieve
agreement with $2+\varepsilon $ expansion, and recent results of functional renormalization group
approach of Ref. \cite{Jakubczyk}. One can see that at $d\sim 5/2$
the exponent $\nu $ is expected to sharply change from the result of $1/N$ expansion with $N\rightarrow N-2$ to
that without the replacement, which seem to correspond to the transition,
predicted in Ref. \cite{2+e_special} and/or crossover discussed recently in Ref. \cite{Jakubczyk}. Following the discussion in pp. (i),
(iii) above, this transition (or crossover) would correspond to the change
from weak to strong spinon confinement. Interestingly, at the dimension $%
d=5/2$ the coefficient $B_{4}$ changes sign, which may be related to the
weak-strong confinement transition, since vanishing of this coefficient
reflects full confinement of spinons. 
%We note that for $d\rightarrow 4$ the
%failure of $1/M$ expansion in the non-compact $CP^{M-1}$ model becomes
%especially spectacular (see Ref. \cite{OurCP}): the lowest order $1/M$
%corrections vanish, while the obtained critical exponents ($\beta =1,$ $\eta
%=2$) are in drastic disagreement with those obtained by other methods.
%Moreover, for $2M<365$ the phase transition in the non-compact $CP^{M-1}$
%model at $d=4-\varepsilon $ was argued to be first order \cite{Failure4-e}.
%The same may be applicable even for $d=3,$ since the critical exponents,
%obtained within the first order of $1/M$ expansion in $CP^{M-1}$ appears to
%be negative (see the discussion in Ref. \cite{OurCP}). Drastic difference of
%these results to the known properties of Heisenberg magnets, described by
%the non-linear sigma model, may also indicate stronger spinon confinement
%for $d\gtrsim 3$ Heisenberg model. From this discussion one observes that
%larger role played by $k^{4-d}$ term posibly goes along with stronger spinon
%confinement.

\section{Conclusions}

In summary, we have obtained the momentum dependence of the self-energy of
spin excitations in the Heisenberg model in the first order $1/N$ expansion.
The obtained dependence contains non-analytic contribution $B_{4}k^{4-d}$ ,
the coefficient $B_{4}$ decreases from positive values at $d\rightarrow 2$
to negative values at $d\rightarrow 4$. We have argued that the non-ananlytical term likely originates 
from the Wilson-Fisher non-Gaussian fixed point. In general dimension $d$, there are also subleading terms 
$B_{2i}k^{2i-(i-1)d}$ with integer values $i>2$, which correspond to the new fixed points, appearing at the dimensions
$d_i=2i/(i-1)$, related also to the relevance of the operators $(\mbox{\boldmath $\phi $}^2)^i$ in the linear sigma model. 

We have shown that the critical exponent $\nu $ in the first order in $1/N$
is determined by two contributions. The first contribution originates from
the exponent $\eta ,$ while the second is proportional to the coefficient $%
B_{4}$ of the non-analytic term. While at $d\rightarrow 2$ the two
contributions almost compensate each other, at $3\lesssim d<4$ the second
contribution dominates and fully determines the value of $1/N$ correction 
to the exponent $\nu $ for $%
d\rightarrow 4.$ The change of sign of the coefficient $B_{4}$ at $d=5/2$ is
associated with the transition (or crossover) from weak to strong spinon
confinement. This is also in line with predicted sharp change of critical
exponents at $d\sim 5/2$ \cite{2+e_special,Jakubczyk}.

Apart from the importance of the obtained results for the interpretation of
numerical and experimental data on Heisenberg magnets, they may have some importance
for itinerant antiferromagnets. Indeed, at half filling the Hubbard model
can be effectively reduced to non-linear sigma model for arbitrary on site
Coulomb repulsion $U,$ see Ref. \cite{Dupuis}. Therefore, one can expect
appearance of the non-analytical terms in the susceptibility in itinerant
half filled antiferromagnets as well. These terms may be rather hard to
obtain from purely fermionic approaches, since they correspond to higher
than four-point fermion interaction vertices. Although numerically correct
value of the critical exponent $\nu $ was obtained previously within the
dynamic vertex approximation (D$\Gamma $A) \cite{OurCrit} and dual fermion
approach \cite{DFcrit} (see also the review \cite{OurRev}), the conclusions
drawn in these studies have to be possibly reexamined in the light of the
results of the present paper, as well as of the most recent D$\Gamma $A
calculations \cite{Toschi2}.

Investigation of the connection between spinon (de)confinement and the
non-analytic term $k^{4-d}$ in the self-energy of spin excitations
represents another important topic for future studies.

\section*{Acknowledgements}

The author acknowledges A. Toschi and G. Rohringer for stimulating
discussions on itinerant magnets, which led to formulation of the problem,
considered in the paper, and P. Jakubczyk for providing the data of Ref. \cite{Jakubczyk}. The work is partly supported by the theme
\textquotedblleft Quant" AAAA-A18-118020190095-4 of Minobrnauki, Russian
Federation.

\begin{widetext}

\appendix

\section{Evaluation of self-energy and mass corrections}

\setcounter{subsection}{0}

\subsection{Dimension $d=3$}

The self-energy is obtained from Eqs. (\ref{CE}) and (\ref{Pi3}) and reads%
\begin{eqnarray}
\Sigma (k) &=&\frac{2}{N}\int \frac{d^{3}\mathbf{q}}{(2\pi )^{3}}\left[ 
\frac{1}{k^{2}+2\mathbf{kq+}q^{2}+m_{0}^{2}}-\frac{1}{q^{2}+m_{0}^{2}}\right]
\frac{4\pi q}{\arctan (q/(2m_{0}))}  \notag \\
&=&\frac{2}{\pi N}\int\limits_{0}^{\Lambda }q^{2}dq\left[ \frac{1}{2kq}\ln 
\frac{(k+q)^{2}+m_{0}^{2}}{(k-q)^{2}+m_{0}^{2}}-\frac{2}{q^{2}+m_{0}^{2}}%
\right] \frac{q}{\arctan (q/(2m_{0}))}
\end{eqnarray}%
By picking out singular contribution 
\begin{equation*}
\Sigma _{1}(k)=\eta k^{2}\ln \left[ \Lambda /(k^{2}+m_{0}^{2})^{1/2}\right]
\end{equation*}%
the intagral can be made convergent. Taking the limit $\Lambda \rightarrow
\infty $ in the remaining part and rescaling the variable of the integration
by $m_{0}$ we obtain Eq. (\ref{SE3}) of the main text. In the following we
denote $\Sigma $ $=\Sigma _{1}(k)+\Sigma _{2}(k)+\Sigma _{3}(k)+\Sigma
_{4}(k),$ where 
\begin{eqnarray*}
\Sigma _{2}(k) &=&\frac{8k^{2}}{9\pi ^{2}}, \\
\Sigma _{3}(k) &=&-4km_{0}/(\pi N),
\end{eqnarray*}%
and $\Sigma _{4}(k)=(2/N)m_{0}^{2}F(k/m_{0})-8k^{2}/(9\pi ^{2})+4km_{0}/(\pi
N)$ is the remaining part, obtained by substracting and adding asymptotic
value of the integrand at large $q$, the function $F(x)$ is given by the Eq.
(\ref{Fxa}). Evaluation of the integrals, which enter Eq. (\ref{Constr1/N})
yields 
\begin{eqnarray}
\int \frac{d^{3}\mathbf{k}}{(2\pi )^{3}}G_{0}(k) &=&\frac{1}{2\pi ^{2}}%
\left( \Lambda -\frac{\pi m_{0}}{2}\right),  \notag \\
\int \frac{d^{3}\mathbf{k}}{(2\pi )^{3}}G_{0}^{2}(k)\Sigma _{1}(k) &=&\frac{%
\eta }{2\pi ^{2}}\left[ \Lambda -\frac{3\pi }{4}m_{0}\ln \frac{\Lambda }{%
m_{0}}\right],  \notag \\
\int \frac{d^{3}\mathbf{k}}{(2\pi )^{3}}G_{0}^{2}(k)\Sigma _{2}(k) &=&\frac{1%
}{2\pi ^{2}}\frac{8\Lambda }{9\pi ^{2}N},  \notag \\
\int \frac{d^{3}\mathbf{k}}{(2\pi )^{3}}G_{0}^{2}(k)\Sigma _{3}(k) &=&-\frac{%
1}{2\pi ^{2}}\frac{4m_{0}}{\pi N}\ln \frac{\Lambda }{m_{0}}  \notag \\
\int \frac{d^{3}\mathbf{k}}{(2\pi )^{3}}G_{0}^{2}(k)\Sigma _{4}(k) &=&\frac{1%
}{2\pi ^{2}}\frac{2m_{0}I}{\pi N},  \notag \\
\int \frac{d^{3}\mathbf{k}}{(2\pi )^{3}}G_{0}^{2}(k) &=&\frac{1}{8\pi m_{0}},
\label{Eqs}
\end{eqnarray}%
where $I=\int_{0}^{\infty }\frac{k^{2}dk}{(k^{2}+1)^{2}}\left( F(k)-\frac{%
4k^{2}}{9\pi }+2k\right) $ is an universal number. Collecting
contributions to the sum rule (\ref{Constr1/N}), which are linear in $%
\Lambda $ or $m_{0}$ \ and do not contain logarithmic terms, we find 
\begin{equation}
m_{0}=\frac{4\pi }{N}\left( 1-\frac{4}{\pi ^{2}N}I\right) \left( \frac{1}{%
t_{c}}-\frac{1}{t}\right),  \label{m03A}
\end{equation}%
where $t_{c}$ is defined according to the Eq. (\ref{tc3}). The remaining contributions to the Eq. (\ref{Constr1/N}) with account
of the last integral in Eqs. (\ref{Eqs}) lead to the mass correction (\ref{dm3}) of
the main text. The resulting Green's function reads%
\begin{equation}
G(k)=\frac{1}{k^{2}+m^{2}+\eta k^{2}\ln \left[ \Lambda
/(k^{2}+m_{0}^{2})^{1/2}\right] +(2m_{0}^{2}/N)F(k/m_{0})}.
\end{equation}%
Collecting log to a power, which is usual in $1/N$ expansion, neglecting
higher order terms in $1/N$ we obtain%
\begin{equation}
G(k)=\frac{1}{\{k^{2}+m^{2}\left[1-\eta \ln \left( \Lambda /(k^{2}+m_{0}^{2})^{1/2}%
\right) \right]+(2m_{0}^{2}/N)F(k/m_{0})\}^{1-\eta /2}},
\end{equation}%
the remaining log contributes to $\nu $ (see below), and the function $%
F(k/m),$ obtained above, describes the non-analytic contribution to the
expression in square brackets. Using $m^2=m_0^2+\delta m^2$ and the expression for the mass correction $\delta m$ (\ref%
{dm3}), we obtain for the Green's fiunction 
\begin{equation}
G(k)=\frac{1}{\{k^{2}+m_{0}^{2}[1+2(\eta +8/(\pi ^{2}N)\ln (\Lambda
/m_{0})]+(2m_{0}^{2}/N)\widetilde{F}(k/m_{0})\}^{1-\eta /2}},
\end{equation}%
where $\widetilde{F}(x)=F(x)-(2/(3\pi ^{2}))\ln [1/(x^{2}+1)].$ Transforming
again logarithmic term into power $m_{0}^{2(\nu -1)},$ denoting $\xi
=m_{0}^{-\nu }$ and neglecting the terms of hihgher order of $1/N\ $we
obtain Eq. (\ref{G3}).

\subsection{Arbitrary $2<d<4$}

In this case we find the polariztion operator%
\begin{eqnarray}
\Pi (q) &=&m_{0}^{d-4}\widetilde{\Pi }(q/m_{0}),  \notag \\
\widetilde{\Pi }(x) &=&\frac{2^{-\frac{d}{2}}\pi A_{d}}{x}\csc \left( \pi
d/2\right) \left( 4+x^{2}\right) ^{\frac{d}{4}-1}  \notag \\
&&\times \left[ \left( \sqrt{4+x^{2}}-x\right) ^{d/2-1}\,_{2}F_{1}\left( 2-%
\frac{d}{2},\frac{d}{2}-1,\frac{d}{2};\frac{1}{2}-\frac{x}{2\sqrt{4+x^{2}}}%
\right) \right.   \notag \\
&&\left. -\left( \sqrt{4+x^{2}}+x\right) ^{d/2-1}\,_{2}F_{1}\left( 2-\frac{d%
}{2},\frac{d}{2}-1,\frac{d}{2};\frac{1}{2}+\frac{x}{2\sqrt{4+x^{2}}}\right) %
\right], 
\end{eqnarray}%
where $_{2}F_{1}(a,b,c;z)\ $is the hypergeometric function, $%
A_{d}=2^{1-d}\pi ^{-d/2}/\Gamma \left( d/2\right) $. For the self-energy we
obtain%
\begin{eqnarray}
\Sigma (k) &=&\frac{4A_{d}}{N}\frac{\Gamma \left( \frac{d}{2}\right) }{\sqrt{%
\pi }\Gamma \left( \frac{d-1}{2}\right) }\int \frac{q^{d-1}dq\sin
^{d-2}\theta d\theta }{\Pi (q)}\left( \frac{1}{k^{2}+2kq\cos \theta \mathbf{+%
}q^{2}+m_{0}^{2}}-\frac{1}{q^{2}+m_{0}^{2}}\right)   \notag \\
&=&\frac{2A_{d}}{N}\int\limits_{0}^{\Lambda }q^{d-1}dq\left[ \frac{%
_{2}F_{1}\left( 1,\frac{d-1}{2},d-1;-\frac{4kq}{(k-q)^{2}+m_{0}^{2}}\right) 
}{(k-q)^{2}+m_{0}^{2}}\,-\frac{1}{q^{2}+m_{0}^{2}}\right] \frac{1}{\Pi (q)}.
\end{eqnarray}%
By substructing and adding asymptotic of integrand at $q\rightarrow\infty$, taking the limit $\Lambda\rightarrow\infty$ in the convergent integral, and rescaling again variable of integration by $m_0$ the result can be put in the from of Eq. (\ref{SE3}) with
\begin{eqnarray}
F_{d}(x) &=&A_{d}\int\limits_{0}^{\infty }q^{d-1}dq\left\{ \left[ \frac{%
_{2}F_{1}\left( 1,\frac{d-1}{2},d-1;-\frac{4xq}{(x-q)^{2}+1}\right) }{%
(x-q)^{2}+1}\,-\frac{1}{q^{2}+1}\right] \frac{1}{\widetilde{\Pi }(q)}\right. 
\notag \\
&&\left. -\frac{2^{d-2}(d-4)\pi ^{\frac{d}{2}-1}\sin \left( \frac{\pi d}{2}%
\right) \Gamma (d-1)}{\Gamma \left( d/2+1\right) }x^{2}q^{-d}\theta (q-\sqrt{%
x^{2}+1})\right\},   \label{Fd}
\end{eqnarray}%
which yields Eq. (\ref{SEd}) of the main text. The lowest order coefficients in the expansion of $F_d(x)$ at large $x$ are given by the Eq. (\ref{B04}) and
\begin{equation*}
B_{2}=\frac{N\eta }{2}\int\limits_{0}^{\infty }tdt\left\{ \frac{d}{4-d}\left[
\frac{t^{2}}{(t-1)^{2}}\,_{2}F_{1}\left( 1,\frac{d-1}{2},d-1;-\frac{4t}{%
(t-1)^{2}}\right) -1\right] -\frac{1}{t^{2}}\theta (t-1)\right\}. 
\end{equation*}%
Evaluating integrals enetering Eq. (\ref{Constr1/N}) we find%
\begin{eqnarray}
\int \frac{d^{d}k}{k^{2}+m_{0}^{2}} &=&A_{d}\left[ \frac{\Lambda ^{d-2}}{d-2}%
+\frac{\pi }{2}\csc \left( \frac{\pi d}{2}\right) m_{0}^{d-2}\right],   \notag
\\
\int \frac{d^{d}k}{(k^{2}+m_{0}^{2})^{2}}k^{2}\ln \frac{\Lambda }{\sqrt{%
k^{2}+m_{0}^{2}}} &=&A_{d}\left[ \frac{\Lambda ^{d-2}}{(d-2)^{2}}+\frac{\pi d%
}{4}\csc \left( \frac{\pi d}{2}\right) m_{0}^{d-2}\ln \frac{\Lambda }{m_{0}}%
+...\right],   \notag \\
\int \frac{k^{2}d^{d}k}{(k^{2}+m_{0}^{2})^{2}} &=&\frac{A_{d}}{d-2}\Lambda
^{d-2},  \notag \\
m_{0}^{d-2}\int \frac{d^{d}k}{(k^{2}+m_{0}^{2})^{2}}k^{4-d}
&=&A_{d}m_{0}^{d-2}\ln \frac{\Lambda }{m_{0}},  \notag \\
\int \frac{d^{d}k}{(k^{2}+m_{0}^{2})^{2}} &=&-A_{d}\frac{\pi (d-2)}{4}\csc
\left( \frac{\pi d}{2}\right) m_{0}^{d-4}.
\end{eqnarray}%
Collecting the terms, proportional to $\Lambda ^{d-2}$ or $m_{0}^{d-2}$ we
find the equation for $m_{0}$%
\begin{equation}
1=tA_{d}\frac{N\Lambda ^{d-2}}{(d-2)}\left[ 1-\frac{\eta }{d-2}-\frac{2}{N}%
B_{2}\right] +NtA_{d}\frac{\pi \csc \left( \pi d/2\right) }{2}m_{0}^{d-2}.
\label{Constr0d}
\end{equation}%
By defining $t_{c}$ according to the Eq. (\ref{tcd}) we obtain%
\begin{equation*}
m_{0}=\left[ -\frac{2}{NA_{d}\pi \csc (\pi d/2)}\left( \frac{1}{t_{c}}-\frac{%
1}{t}\right) \right] ^{1/(d-2)}.
\end{equation*}%
The correction $\delta m^{2}$ is obtained then straightforwardly from the
remaining terms in the sum rule (\ref{Constr1/N}) and given by the Eq. (\ref%
{dmd}). Repeating the calculation of the Green's function similarly to $d=3$
case we find%
\begin{equation}
G(k)=\frac{1}{\{k^{2}+m_{0}^{2}[1+(2\eta /(d-2)+(8B_{4}/\pi N)\sin \left(
\pi d/2\right) )/(d-2)\ln (\Lambda /m_{0})]+(2m_{0}^{2}/N)\widetilde{F}%
_{d}(k/m_{0})\}^{1-\eta /2}}
\end{equation}%
where $\widetilde{F}_{d}(x)=F_{d}(x)-(N\eta /4)\ln [1/(x^{2}+1)].$ After
transforming logarithm into power we obtain again the result (\ref{G3}) with 
$\xi =m_{0}^{-\nu (d-2)}\propto (t-t_{c})^{-\nu },$ $\nu $ is given by the
Eq. (\ref{nud}). 

\subsection{Dimension $d=4$}

For completeness, let us also present some results in four dimensions. Performing integration in Eq. (\ref{Pi}), we obtain
\begin{equation}
\Pi (q)=\frac{1}{8\pi ^{2}}\left[ \ln \left( \frac{\Lambda }{m_0}\right) -%
\frac{\sqrt{4m_0^{2}+q^{2}}}{2q}\tanh ^{-1}\left( \frac{q\sqrt{4m_0^{2}+q^{2}}}{%
2m_0^{2}+q^{2}}\right) +\frac{1}{2}\right]. 
\end{equation}
The corresponding contribution to the self-energy reads
\begin{equation}
\Sigma (k)=\frac{1}{{4\pi ^{2}N}}\int q^{3}dq \frac{1}{{\Pi (q)}}\left[ \frac{%
k^{2}+q^{2}+m_0^{2}-\sqrt{k^{4}+( q^{2}+m_0^{2})
^{2}-2k^{2}(q^{2}-m_0^{2})}}{2k^{2}q^{2}}-\frac{1}{q^{2}+m_0^{2}}\right]. 
\end{equation}
After evaluating the integral in the limit $k\gg m_0$ and neglecting the terms of the order of $k^2/l$ and $m_0^2/l$, where $l=\ln(\Lambda/k)$, $\ln(\Lambda/m_0)$, or $\ln(k/m_0)$, we obtain
\begin{equation}
\Sigma (k)=-\frac{6m_0^2}{N}\left[ \frac{2\ln (k/m_0)}{1+2\ln (\Lambda /k)}-\ln \left( \frac{2\ln (\Lambda /m_0)}{1+2\ln
(\Lambda /k)}\right) \right]. 
\end{equation}
We note that the coefficient in front of the square bracket is equal to $2B_4(d\rightarrow 4)/N$. Performing integrations in Eq. (\ref{Constr1/N}), we obtain with logarithmic accuracy%
\begin{eqnarray}
\int \frac{d^{4}k}{(2\pi )^{4}}\frac{1}{(k^{2}+m_{0}^{2})} &=&\frac{1}{8\pi
^{2}}\left( \frac{\Lambda ^{2}}{2}-m_{0}^{2}\ln \frac{\Lambda }{m_{0}}\right),\notag
\\
\int \frac{d^{4}k}{(2\pi )^{4}}\frac{1}{(k^{2}+m_{0}^{2})^{2}}\Sigma (k) &=&%
\frac{3m_{0}^{2}}{4\pi ^{2}N}\ln \frac{\Lambda }{m_{0}}\left( 2-\ln \ln 
\frac{\Lambda }{m_{0}}\right),\notag  \\
\int \frac{d^{4}k}{(2\pi )^{4}}\frac{1}{(k^{2}+m_{0}^{2})^{2}} &=&\frac{1}{%
8\pi ^{2}}\ln \frac{\Lambda }{m_{0}}.
\end{eqnarray}%
Putting $m_0=\delta m=0$ we find the critical temperature $t_{c}=16\pi ^{2}/(\Lambda ^{2}N)$. Absorbing $\ln(\Lambda/m_0)$ contributions into the bare mass $m_0$ we find 
%Constraint equation%
%\begin{eqnarray}
%\frac{Nt}{8\pi ^{2}}\left[ \frac{\Lambda ^{2}}{2}-m_{0}^{2}\ln \frac{\Lambda 
%}{m_{0}}\right] -\frac{3m_{0}^{2}}{4\pi ^{2}}t\left[ \ln \frac{\Lambda }{%
%m_{0}}\left( 2-\ln \ln \frac{\Lambda }{m_{0}}\right) \right] -Nt\frac{\delta
%m^{2}}{8\pi ^{2}}\ln \frac{\Lambda }{m_{0}} &=&1 \\
%\frac{Nt}{8\pi ^{2}}\left[ \frac{\Lambda ^{2}}{2}-m_{0}^{2}\left( 1+\frac{12%
%}{N}\right) \ln \frac{\Lambda }{m_{0}}\right]  &=&1
%\end{eqnarray}%
\begin{equation}
m_{0}^{2}\ln \frac{\Lambda }{m_{0}}=\frac{8\pi ^{2}}{N+12}\left( \frac{1}{%
t_{c}}-\frac{1}{t}\right). \label{m04d}
\end{equation}%
Finally, remaining contributions to the Eq. (\ref{Constr1/N}) yield
\begin{equation}
\delta m^{2}=\frac{6m_{0}^{2}}{N}\ln \ln \frac{\Lambda }{m_{0}}.\label{dm4d}
\end{equation}%
In Eqs. (\ref{m04d}) and (\ref{dm4d}) we recognize the zeroth- and first order terms in $1/N$ expansion of the one loop renormalization group result (see, e.g., Ref. \cite{Amit}) $m^2 \propto(t-t_c)/\ln^{(N+2)/(N+8)}(\Lambda/m)$.
%\begin{equation}
%\Sigma (k)+\delta m^{2}=\frac{6m_{0}^{2}}{N}\left[ \ln \frac{\ln (\Lambda
%/m)^{2}}{\ln (\Lambda /k)}-\frac{\ln (k/m)}{\ln (\Lambda /k)}\right] 
%\end{equation}

\end{widetext}

\end{document}